\documentclass[submission]{eptcs}
\providecommand{\event}{Refinement Workshop 2013} 
\usepackage{breakurl}             

\usepackage{bsymb,url}

\usepackage{graphicx}

\title{Modelling and Refinement in CODA}
\author{Michael Butler \quad John Colley\quad Andy Edmunds \quad Colin Snook
\institute{Electronics and Computer Science, University of Southampton, UK}
\and
Neil Evans \quad Neil Grant \quad Helen Marshall
\institute{AWE, UK}
}
\def\titlerunning{Modelling and Refinement in CODA}
\def\authorrunning{Butler Et Al}
\date{28 March 2013}
\begin{document}
\maketitle

\begin{abstract}
This paper provides an overview of the CODA framework for modelling and refinement of component-based embedded systems. CODA is an extension of Event-B and UML-B and is supported by a plug-in for the Rodin toolset. CODA augments Event-B with constructs for component-based modelling including components, communications ports, port connectors, timed communications and timing triggers. Component behaviour is specified through a combination of UML-B state machines and Event-B. CODA communications and timing are given an Event-B semantics through translation rules. Refinement is based on Event-B refinement and allows layered construction of CODA models in a consistent way.
\end{abstract}

\section{Introduction}

Simulation-based modelling methods for embedded systems are typically structured in terms of communicating model components as this closely reflects their design structure.  
In addition, diagrammatic languages such as UML (Unified Modelling Language)~\cite{UML} provide intuitive notations for representing both architectural structure (e.g., component diagrams) and behaviour (e.g., state machine diagrams) and these find 
acceptance in industrial practice.  
While simulation tools play an indispensable tool in system verification, these approaches typically lack two important development concepts: layering of models at multiple abstraction levels (i.e., model refinement) and formal verification.  CODA (Co-Design Architecture) is a component-based modelling framework that aims to combine the advantages of
component-based graphical modelling and simulation with refinement and formal verification. These aims fulfil AWE's need for an engineer-friendly environment for systems development with a formal underpinning.

AWE is particularly focused on embedded systems that are combinations of software and hardware. The original proposal for CODA was inspired by Sandia National Laboratories' simulation environment called Orchestra \cite{DBLP:conf/wotug/Wickstrom07} within which system models can be refined and decomposed into a collection of communicating components, representing as software, hardware or  mechanical devices. While Orchestra does support modelling at different abstraction levels, it is a simulation-based approach and does not have a formal definition of refinement nor does it include formal 
verification.

Rather than starting from scratch, we have defined CODA as an extension to the existing Event-B formal approach~\cite{abrial2010modeling}. 
In Event-B  a system is modelled in terms of  state variables and guarded events that alter that state. Central to Event-B is the notion of refinement that allows essential properties to be expressed at a very abstract (hence simple and clear) level and then progressive refinements allow more and more detail to be added until the full detail of the system has been described. At each refinement the consistency of the model has to be proven including the correctness of the refinement (i.e. that no new traces have been introduced and that the refined state has equivalence with the abstract state). Defining CODA as an extension of Event-B is in
the spirit of UML-B which augments Event-B with specialisations of UML entities, including UML class diagrams and state machine diagrams~\cite{snook2006ubf}.  UML-B includes notions of refinement that correspond to natural extensions of UML-like diagrams: class diagrams are refined through class extension and class addition mechanisms while state machines are refined by added nested substates within more abstract states~\cite{Said:2009:LTS:1693345.1693393}.

In CODA we inherit the UML-B state machine construct and specialise it for our purpose. We also add component diagrams with constructs that are influenced by the Orchestra approach.  A component diagram 
consists of a collection of individual components. Following the Orchestra approach, components interact through asynchronous timed channels.  The behaviour of a component is defined through one or more state machines. Notational extensions are provided with operations for reading and writing to timed channels, as well
as constructs for specifying timing-based `wake-ups'.  It is also possible to define some component behaviour 
through textual Event-B guards and actions directly.  In CODA we typically start with an abstract model that has a small number of components and some simple state machines that capture essential properties of a system.
We use three main forms of structural refinement: addition of new components, addition of state machines, and refinement of state machines through addition of nested state machines.

In Event-B, an abstract model comprises a \emph{machine} that specifies the high-level behaviour and a \emph{context}, made up of sets, constants and their properties, that represents the type environment for the high-level machine.  The machine is represented as a set of \emph{state variables}, $v$ and a set of events, \emph{guarded atomic actions}, which modify the state. If more than one action is enabled, then one is chosen non-deterministically for \emph{execution}. Event-B defines proof obligations to ensure that events preserve  \emph{invariants} on the variables.  A more concrete representation of the machine may then be created which refines the abstract machine, and the abstract context may be extended to support the types required by the refinement.  \emph{Gluing invariants} are used to verify that the concrete machine is a correct \emph{refinement}: any behaviour of the concrete machine must satisfy the abstract behaviour.  Gluing invariants give rise to proof obligations for pairs of abstract and corresponding concrete events. 

We have developed tool support for CODA based on the Rodin toolset~\cite{AbrialBHHMV10} .
The Rodin toolset is an extensible environment for modelling with Event-B. It includes automatic tools to generate proof obligations of the model’s consistency and provers that attempt to automatically discharge these obligations. ProB ~\cite{eps262886} is a model checker and animator that is available as an extension to the Rodin toolset.
The UML-B plug-in supports modelling and refinement of class diagrams and state machines and translates models into Event-B for animation, model checking and proof. 
 We have developed  a prototype framework for the CODA approach that supports modelling at different refinement levels  and is integrated as a plug-in within the Rodin environment.  This enables animation, model checking and proof of CODA models. Rodin, ProB and UML-B are important ingredients that we have built on in order to achieve the CODA tool.

Since our work was inspired by Orchesra, we provide an overview of Orchestra in the next section.  
We then proceed with an overview of the CODA modelling language (Section~3) followed by an outline of how CODA models are mapped to Event-B (Section~4).
Section~5 provides an outline of refinement in CODA an Section~6 outlines the application of the CODA approach to a washing macchine case study.
In Section~7 refinement in CODA is extended to cover refinement of  inputs and outputs.
Section~8 outlines the role of model checking and simulation in CODA development and Section~9 concludes the paper.

\section{Background on Orchestra}
 
\begin{figure}[!htb]
  \centering
  \includegraphics[width=14cm]{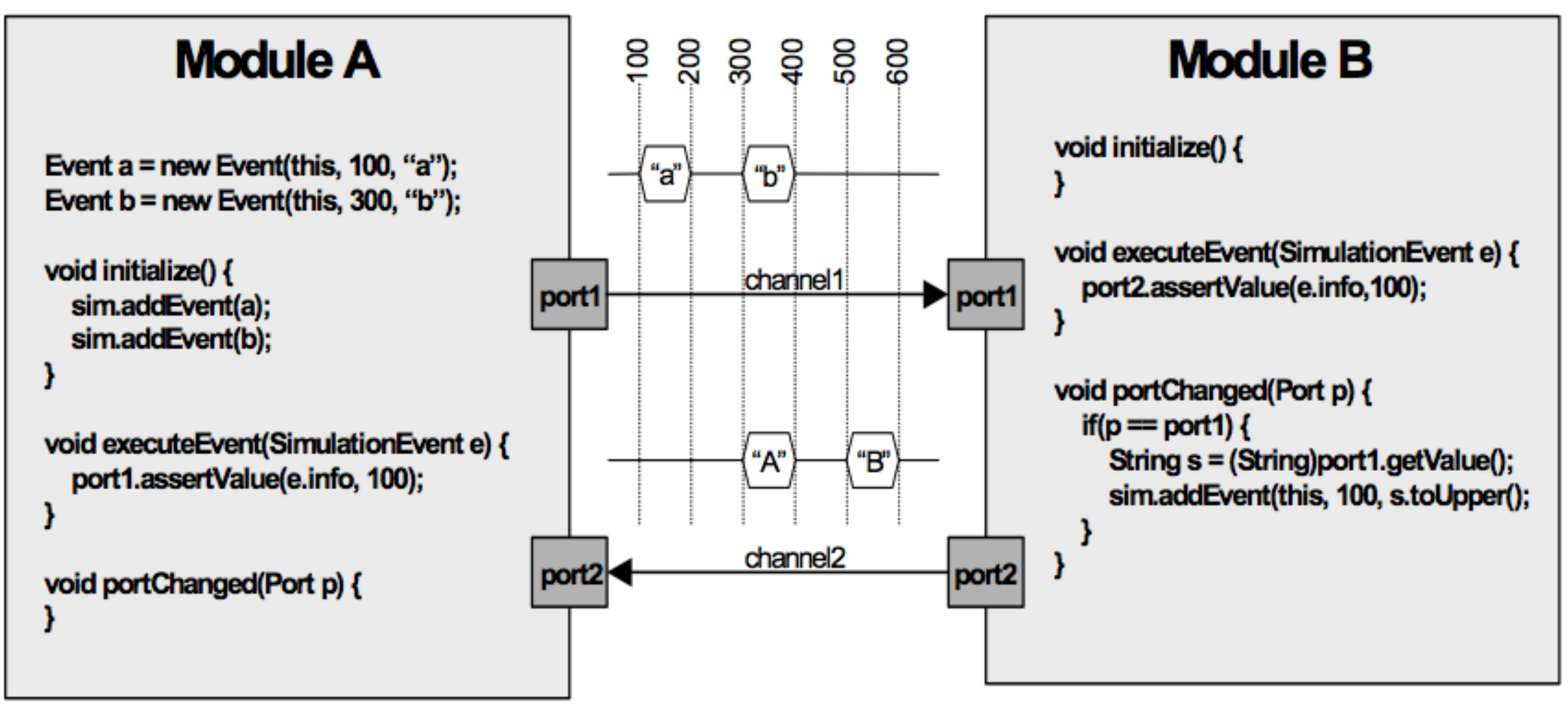}
  \caption{The Orchestra API}
  \label{Figure:figOrchestraAPI}
\end{figure}

Orchestra is a simulation environment for embedded systems \cite{DBLP:conf/wotug/Wickstrom07} within which system models can be refined and decomposed into a collection of communicating components, modelled as software, hardware (including emulated hardware) or as mechanical devices.  The notion of refinement in Orchestra, however, is not formal and is not supported by any formal method. 
Orchestra provides a discrete event simulation environment which is based on the established simulation technology used by industrial Verilog \cite{thomas2002vrh}, VHDL \cite{perry1994v}  and SystemC  \cite{mueller2001sss} simulators. The essence of the Orchestra environment is described in Figure~\ref{Figure:figOrchestraAPI}  which is taken from \cite{DBLP:conf/wotug/Wickstrom07} .

Orchestra \emph{simulation modules} communicate via input and output \emph{ports} which are connected using channels.  A module may also communicate directly with another module using a \emph{method call}.
The Orchestra API provides an object-oriented interface that allows synchronisation and communication using the following fundamental calls.

The Orchestra $sim.addEvent(e)$
method call encapsulates the event \emph{e} and the \emph{simulation time} at which a \emph{call-back} method, \emph{executeEvent}, provided by the model developer, will be invoked by the simulator kernel. Within the call-back method, the developer can \emph{get} or \emph{assert} values on ports.

The $porto.assertValue(e.info,duration)$
 method call will assert a \emph{value}, \emph{e.info}, on the channel connected to the output port \emph{porto}, and this value will be made available to any \emph{input port} connected to this channel after a delay of  value \emph{duration} via a \emph{call-back} method, \emph{portChanged}, provided by the model developer. If a module is connected to more than one channel via input ports and the value changes on more than one input port simultaneously, the call-back is only invoked once.

The $porti.getValue()$
 method call returns the value on the input port \emph{porti}.

\section{Overview of the CODA modelling language}

\begin{figure}[t]
  \centering
  \includegraphics[width=12cm]{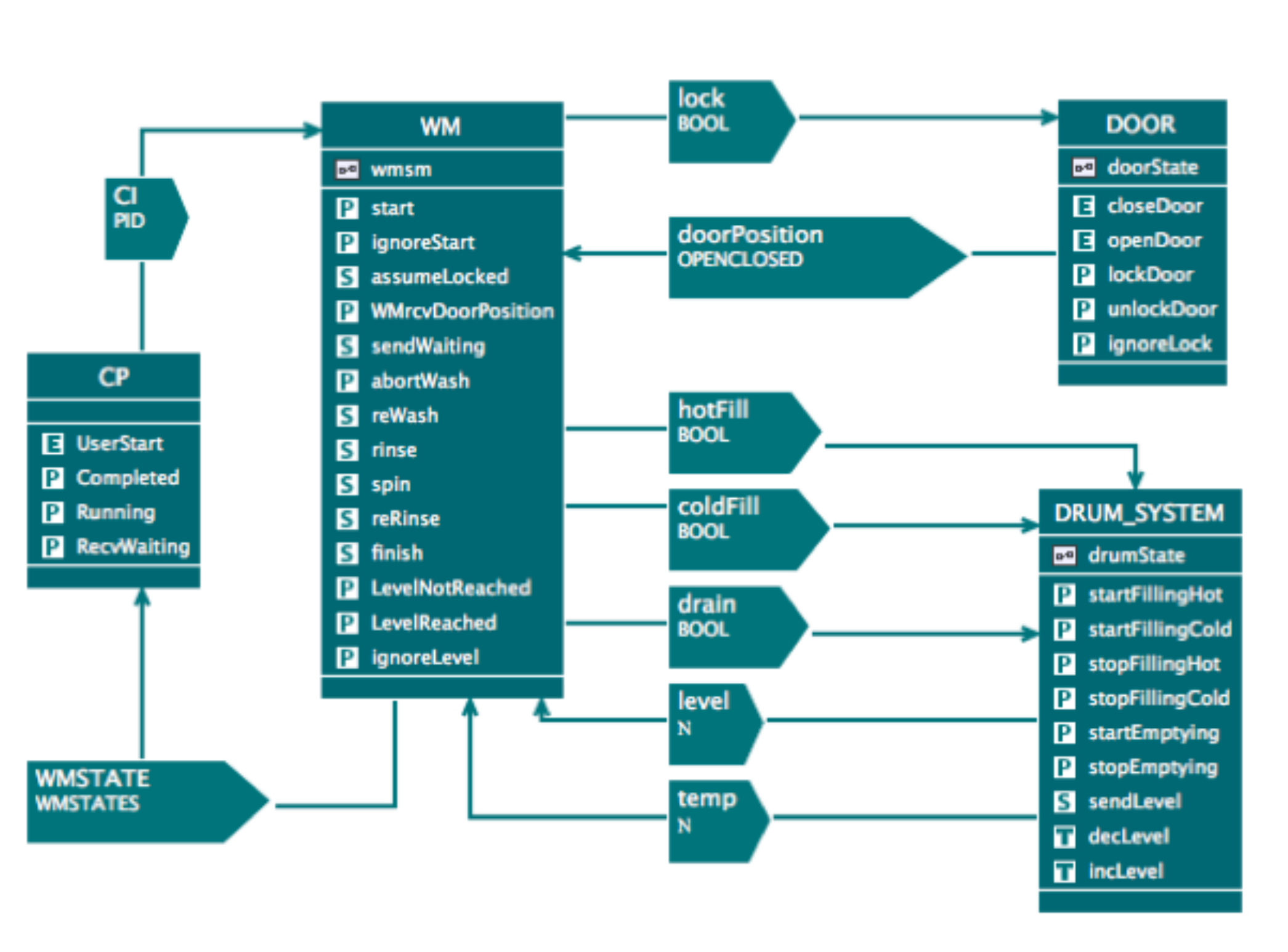}
  \caption{Component Diagram for a Washing Machine}
  \label{Figure:WM4}
\end{figure}

Figure~\ref{Figure:WM4} shows a CODA component diagram for a washing machine system. The model consists of four components: a control panel ($CP$), a door system, a drum system and washing machine controller ($WM$).  These components are connected via typed asynchronous channels.  For example, the $lock$ channel is used to send boolean messages from the $WM$ component to the $DOOR$ component while the $level$ channel is used to send water level values from the $DRUM\_SYSTEM$ to  $WM$.  Model specific types may be introduced in Event-B contexts and used to define channel types, such as the $PID$ type used in the $CI$ channel from $CP$ to $WM$.

Components may contain state machines and operations and these are listed within the component on the diagram.  For example $WM$ includes a state machine called $wmsm$ (immediately below the $WM$ name in the $WM$ box. The $WM$ component also contains a list of operations such as $start$, $ignoreStart$, $assumeLocked$ etc. 

If a component is connected to the input side of a channel, then operations of that component may perform \textit{port-send} actions on that channel.  
A port-send action represents the action of sending data over a connector and can be added to any operation in a component that has an outgoing connection to the connector. When a message is sent, it is specified to be delivered at some later time, defined by a $delay$ value.  
A component  connected to the output side of a channel has special operations, called \textit{port-wake} operations, that are triggered when a message arrives at that component from the channel.  
 A port-wake operation is needed in the receiving component of the connector in order to respond to the receipt of data on the connector.  A port-wake operation is always associated with exactly one channel. For example, the $startFillingHot$ port-wake operation of the $DRUM\_SYSTEM$ is associated with the $hotFill$ channel meaning when a value becomes available at the receiving end of that channel, then the $startFillingHot$ operation is executed.

A component can also set an internal wake-up through a \textit{self-wake} setting action.  Setting a self-wake causes a {self-wake} operation to be triggered at a later  time. 
Components may also have operations that represent environment events, internal transitions of state machines and method calls used for modelling software method calls.  

Component operations can be one of five types as indicated by the letters $P, S, E, T, M$:
\begin{description}
\item[P]  Port-wake operation which is triggered by arrival of a message from a channel at the specified time.
\item[S]  Self-wake operation which is triggered by the expiry of a self-wake.
\item[E] Environment operation that represent an external stimulus from the environment.
\item[T] Transition operation that represents  transitions of a state machine.
\item[M] Method operation.
\end{description}

\paragraph{Timing:} Similar to Orchestra, CODA includes a timing model based on a global dicrete clock. 
A port-send operation sets the future time at which the corresponding port-wake operation will be triggered through a delay argument.
Similarly self-wake  operations are triggered to occur at a specific future time when the self-wake is set.
Operations and transitions may be \textit{synchronised}, meaning they occur at most once per clock cycle, or \textit{unsynchronised}, meaning they may occur multiple times per cycle.
Port wakes, self wakes and methods are synchronised.
Environment operations are not synchronised with the clock.  We assume that a finite number of environment operations may occur within a clock cycle.  In practice, environment operations are assumed to occur more slowly than the system clock.
State machine transitions may be synchronised or unsynchronised.  We need to ensure that only a finite number of unsynchronised transitions may occur per clock cycle.  This can be achieved through a convergence proof, i.e., exhibiting a variant that is decreased by unsynchronised transitions.

From the beginning, CODA was designed to support a state machine based development and verification method.  CODA re-uses the state machine modelling and refinement capabilities of UML-B unchanged and is therefore able to leverage any developments in this area.
The CODA Component View introduces a graphically-based level of abstraction.  Our initial work on CODA using just state machines confirmed AWE's need for a Component View as it fits in with engineering practice.  They demonstrated that a state machine abstraction was not sufficient in itself to enable CODA modelling at an appropriately high level of abstraction. Even though the facility for delayed communication was already in place, it was difficult for the user to introduce this communication at the state machine level.  The Component View encapsulates the component and connections in a clear way and enables the user easily to annotate the component diagrams with timing information at a high level of abstraction.  The Component View and the State Machine View sit side by side to enable efficient development and refinement of CODA models.

\begin{figure}[t]
  \centering
  \includegraphics[width=8cm]{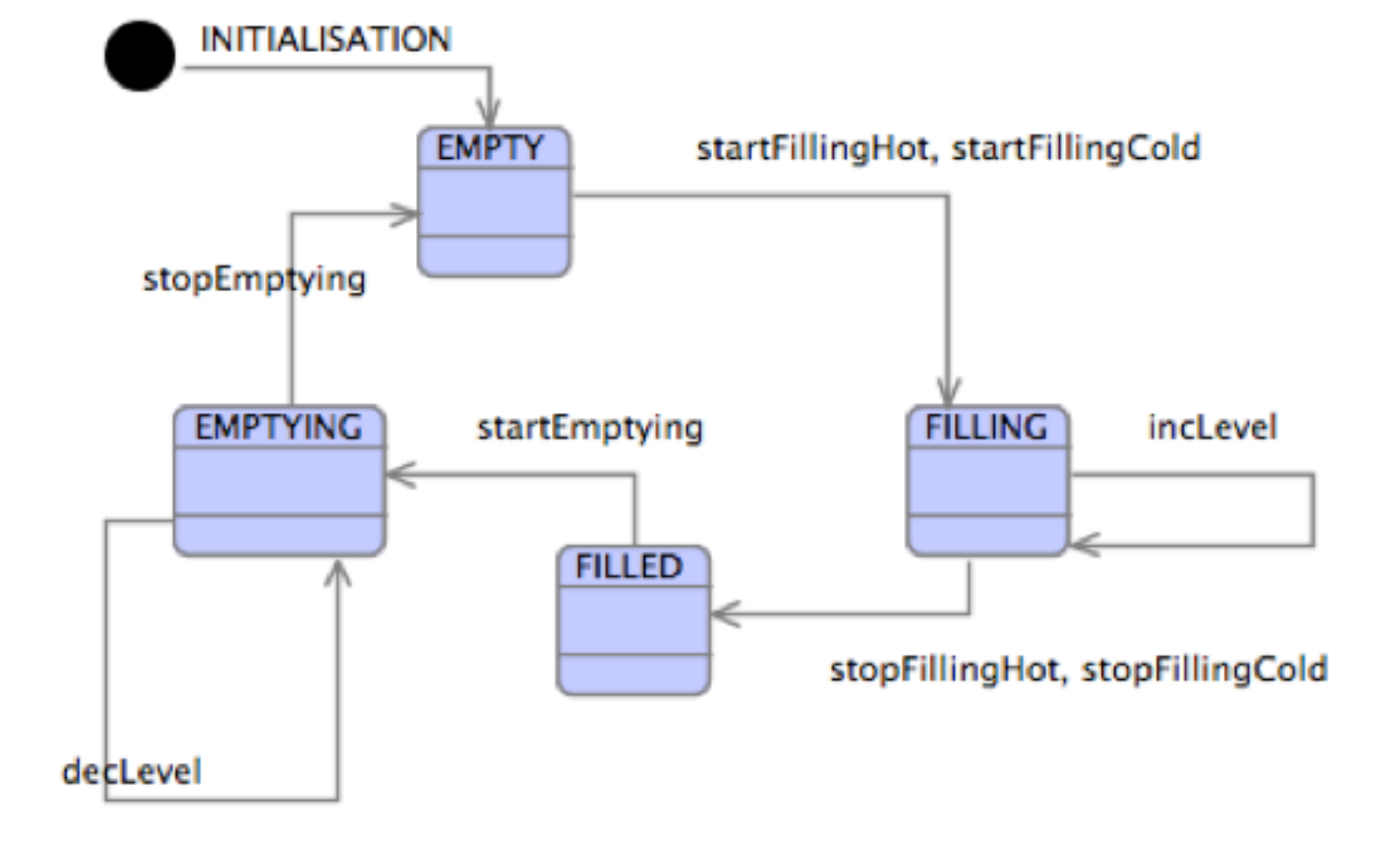}
  \caption{State Machine for Drum System Component}
  \label{Figure:Drum1}
\end{figure}

Figure~\ref{Figure:Drum1} shows the $drumState$ state machine that is contained with the $DRUM\_SYSTEM$ component of the washing machine.   Operations may be associated with a transition of a state machine, meaning that state transition is synchronised with execution of the operation.  For example, the $startFillingHot$ port-wake operation of the $DRUM\_SYSTEM$ is associated with the transition from the $EMPTY$ state to the $FILLING$ state of the $drumstate$ state machine.  This means that when the $DRUM\_SYSTEM$ receives a message on a channel (in this case the $hotFill$ channel) it makes the transition from the $EMPTY$ state to the $FILLING$ state.

To model software behaviour, method call actions can be added to operations and corresponding method operations must be added to service the calls. A method call action immediately enables a particular method operation, which must then complete within the same clock cycle.
External operations represent events that occur in the environment of a controller. The timing of these events is uncontrolled and therefore not synchronised to the clock. These events may perform any of the operation actions (port-wake, self-wake, method calls) described above.
Two kinds of state-machine are available. Asynchronous state-machines are not linked to the clock whereas synchronous state-machines are tightly linked to the clock so that, while enabled, only one transition may be taken on each clock tick.

Note that the CODA model shown in Figure~\ref{Figure:WM4} represents the fourth level of a refinement chain.  In Section~\ref{sec:wm}, we will see how this is arrived at from a more abstract model with a simpler structure.

\section{Mapping CODA models to Event-B}

CODA models are given a formal semantics through translation to an underlying Event-B model.
State machines are mapped to an underlying Event-B model in the standard UML-B way as described in~\cite{Said:2009:LTS:1693345.1693393}.  In this paper we focus on the representation of the port- and self-wake operations. 
Component operations contribute guards and actions to an event in the underlying Event-B model. Operations may have ordinary guards and actions expressed in the Event-B notation but they will also have special kinds of guards and actions associated with port-wakes, self-wakes and method calls.

A  connector is modelled in Event-B as a partial mapping from discrete time points to values of an appropriate type as follows:
$$ 
	connector ~\in~ \nat \pfun type
$$
Thus, if we have $t\mapsto v\in connector$, then value $v$ is received from that connector at time $t$ triggering the associated port-wake operation to be invoked.
Use of timing constructs gives rise to a variable in the underlying Event-B model representing the current time.
A send action is of the form 
$$
	send(connector,v,delay)
$$
This specifies that value $v$ should be sent on the connector to be received at exactly $delay$ time units in the future.
This is mapped to the following underlying Event-B action:
$$
connector (current\_time+delay) ~:=~ value
$$

When a value is received from a connector, it should be the most recent value on the connector.
The mapping from discrete time points to values is sparse so the following guard is used to specify the most recently available value on the connector (that must not be in the future):
$$
	connector(max( \{~ t ~|~ t \in dom(connector) ~\land~ t \leq current\_time\})) = value
$$

For synchronisation, port-wake operations are grouped according to the combination of incoming connectors that they respond to. Therefore if two port-wake operations in the same component have port-wake properties on the same group of connectors they will use the same synchronisation flag and hence exactly one of them will respond to the simultaneous arrival of data on that group of connectors. Which one does so may be controlled via other guards such as particular values arriving on the connectors.  If more than one event should occur as a result of a part-wake, then the port-wake calls a method.

Self-wakes are modelled in a similar way to connectors.  
A component can be scheduled to wake at some time in the future by adding to a queue of wake events. This results in a queue variable in the underlying Event-B, one for each component, to contain the scheduled event times specified as follows:
$$ 
	component\_wakeup ~\in~ \nat \pfun WakeKind
$$
The range type here is $WakeKind$ to allow for different kinds of wake event, for example, a collection of interrupt priorities. (At the moment only one kind is supported.)
In CODA a wake-up is set using an action of the following form:
$$
	self\_wake(component,kind,delay)
$$
The underlying Event-B representation for this is as follows:
$$
component\_wakeup (current\_time+delay) ~:=~ kind
$$

When a component wake up event is reached, all of the component's self-wake operations are enabled subject to their other guards. 
This time based enabling of self-wake operations is represented with the following guard in the underlying Event-B:
$$ 
	current\_time \in dom(component\_wakeup)
$$

As well as requiring the current time variable, the underlying Event-B has an event for advancing time.
There should be no pending connector receives nor pending wakes at the current time if the time is progressed, thus the event for advancing time has guards as follows for each connector and component:
$$
	current\_time \not\in ( dom(connector_i) )
$$
$$
current\_time \not\in dom(comp_{i}\_wakeup)
$$

\section{Refinement in CODA}

In the CODA method, modelling begins with an abstract, un-timed model of the system, together with its environment, as a single component which represents the abstract specification of the system together with its interactions with the environment.

In the first refinement, a feature of the system is singled out and the rest of the system/environment is left at the abstract level.  The top-level abstract component is therefore refined into two components, one that represents the feature to be modelled in more detail and one that represents the rest of the system/environment, with timed communication established between the components by means of one or more connections.  
Typically the introduction of communication events in a refinement requires the refinement of state machines through the introduction of nested states.  It must now be established that this two-component model is a correct refinement of the abstract single component model.

In subsequent refinements,  features of the system are singled out from the abstract component for attention;  after the second refinement, two components now communicate via timed connections  with what remains of the abstract system/environment component.  The component chosen may be a target hardware/software component, a mechanical component or a system environment component such as a control surface.

The modelling of time in CODA ensures that this refinement can be verified formally;  as outlined in the previous section, time is modelled as a natural number and always advances by a single tick, the finest possible granularity of activity within the CODA model as it is refined.  The CODA modeller is able to specify delayed communication at a high-level of abstraction in terms of \emph{port sends}, \emph{port wakes} and \emph{self wakes}.  These abstractions have been developed in a way that does not inhibit refinement proof;  starting with an un-timed abstraction, it is possible to introduce more and more detailed timing information in successive refinements.

Systematic refinement continues until every component of the system and its environment has been modelled.  The refinement method does not restrict the modeller and supports back-tracking to an abstract model to make changes.  At this stage, the high-level CODA model is complete. 

Proof of the correctness of the CODA refinements is undertaken at the level of the underlying Event-B.
On the case studies that we have undertaken, we have found that, at each stage of this refinement process, it is possible within CODA either to prove that each refinement step is  correct, or to reveal unforeseen problems that require re-modelling.

\section{Example} \label{sec:wm}

\subsection{Abstract Model}

\begin{figure}[!htb]
  \centering
  \includegraphics[width=11cm]{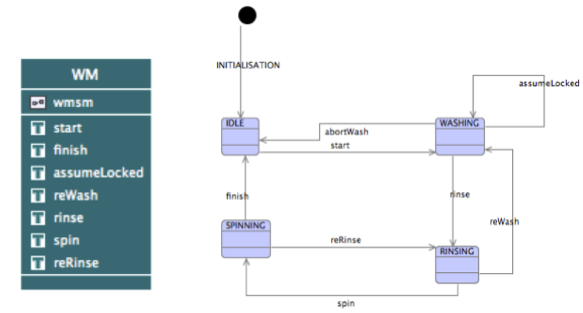}
  \caption{Abstract Component Diagram for Washing Machine}
  \label{Figure:WM1}
\end{figure}

The modelling process begins by describing a single, abstract state machine $wmsm$ that represents the washing machine together with its environment. Four states represent the modes of the system: \textit{IDLE}, \textit{WASHING}, \textit{RINSING} and \textit{SPINNING} and seven transitions represent how the system modes evolve. A component \textit{WM} is also introduced to represent the complete system. It contains the state-machine and owns operations that link to the transitions in the state-machine. The top-level component and state machine are shown in Figure~\ref{Figure:WM1}.

This system-level state machine is un-timed and non-deterministic. For instance, when the system is in state \textit{RINSING}, the system will immediately move to either state \textit{SPINNING} or \textit{WASHING} non-deterministically. The state machine represents all the possible mode traces of the system.

Before proceeding, we animate the system-level state machine to validate that states and transitions correctly represent the system-level view of the washing machine. This is a validation process requiring subjective evaluation of the model against system requirements. Graphical animation of the state machines is supported by the UML-B plug-in.
In addition to animation,  proof and model checking  can be applied to the underlying Event-B model in the usual way with Rodin and ProB.

\subsection{First Refinement}

\begin{figure}[!htb]
  \centering
  \includegraphics[width=10cm]{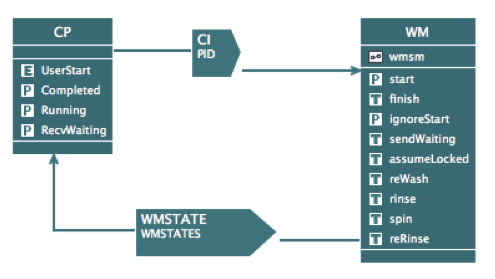}
  \caption{First Refinement: Introduce the Control Panel}
  \label{Figure:WM2}
\end{figure}

The single component and abstract state machine is now refined into a system comprising two components as shown in Figure~\ref{Figure:WM2}. The first component is the Control Panel and the second the abstract washing machine sub-system. Two connectors enable communication between the two components. The first connector, CI, is used to pass the Washing Program ID ($PID$) to the washing machine sub-system and the second connector, $WMSTATE$, passes the status of the sub-system back to the Control Panel to be displayed. The state machine is unchanged except for the addition of a self transition on state $IDLE$ which constrains the $sendWaiting$ operation so that it only sends the waiting status over the $WMSTATE$ connector while the washing machine is idle.

The external operation, $UserStart$, in component $CP$ represents the user starting the wash by passing the selected wash program, using a port-send action on connector $CI$ to the washing machine sub-system. 
A corresponding port-wake operation, $start$, in the washing machine sub-system receives the program ID that will, in a subsequent refinement, be decoded to control the wash. 
A further port-wake operation, $ignoreStart$, manages inadvertent start requests from the user. Note that this is necessary due to a design decision not to constrain the sending of start messages from $CP$. If $WM$ is not in a state to respond to the start an explicit $ignoreStart$ is needed to avoid the system deadlocking.
When the washing machine sub-system receives the $pid$, it responds with a port-send action on connector $WMSTATE$ to inform the Control Panel that the washing machine is now $RUNNING$.
The Control Panel receives the message from the washing machine sub-system with the port-wake operation $Running$ so that this information can be displayed to the washing machine user.

\subsection{Second Refinement}

\begin{figure}[!htb]
  \centering
  \includegraphics[width=10cm]{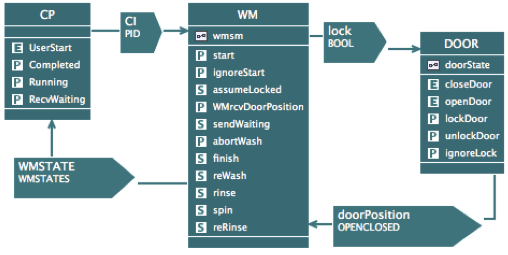}
  \caption{Second Refinement: Introduce the Door Component}
  \label{Figure:WM3}
\end{figure}

The washing machine sub-system component is now further refined, as shown in Figure~\ref{Figure:WM3}, into two components, the $DOOR$ sub-system and an abstract component, $WM$, that represents the rest of the washing machine sub-system. Two connectors enable communication between these two components. The first, $lock$, passes a Boolean signal to the $DOOR$ sub-system to lock the door. The second, $doorPosition$, informs the Washing Machine sub-system when the door is opened or closed.

Note that the DOOR component has two external operations, $closeDoor$ and $openDoor$, which represent the interaction of the user with the door. Care is needed in this refinement to ensure that the system cannot get into an unsafe state; the door should always be locked when the washing machine is washing, rinsing or spinning so that the user cannot inadvertently open the door and release potentially very hot water.

\begin{figure}[!htb]
  \centering
  \includegraphics[width=10cm]{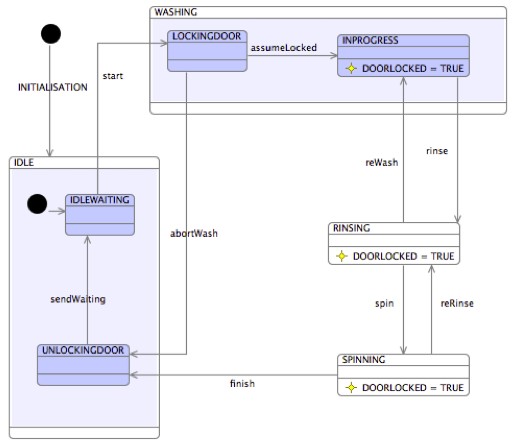}
  \caption{Refined State Machine of $WM$ in the Second Refinement}
  \label{Figure:WM3sm}
\end{figure}

The state-machine for the washing machine is refined to split the $WASHING$ state into sub-states $LOCKINGDOOR$ and $INPROGRESS$ and $IDLE$ into U$NLOCKINGDOOR$ and $IDLEWAITING$ (Figure~\ref{Figure:WM3sm}). This is necessary to accommodate the new transitions concerned with locking and unlocking the door.  Comparing the state machine of Figure~\ref{Figure:WM1} with the state machine of Figure~\ref{Figure:WM3sm}, we can see the introduction of nested states in the $IDLE$ and $WASHING$ states.

An invariant, 
$$DOORLOCKED = TRUE,$$
 is introduced in the sub-system state machine for states $INPROGRESS$, $RINSING$ and $SPINNING$.

\begin{figure}[!htb]
  \centering
  \includegraphics[width=9cm]{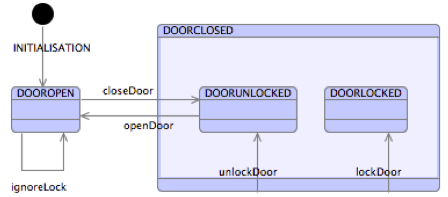}
  \caption{New State Machine for the Door Component in the Second Refinement}
  \label{Figure:WM3smDoor}
\end{figure}

A new state machine is introduced for the $DOOR$ component as shown in Figure~\ref{Figure:WM3smDoor}.
The door may be open ($DOOROPEN$) in which case any instructions to lock the door are ignored ($ignoreLock$) or it may be closed ($DOORCLOSED$). When the door is closed it may be unlocked ($DOORUNLOCKED$) or locked ($DOORLOCKED$). Note that the transitions $unlockDoor$ and $lockDoor$ are drawn with the superstate $DOORCLOSED$ as their source indicating that they can fire irrespective of whether the door is locked or not.

The washing machine sub-system sends a message via the $lock$ connector to the door sub-system to lock the door if it has received a message from the door via the $doorPosition$ connector indicating that the door is closed. The washing machine sub-system then initiates a self-wake, delayed by 3 time units. If the door is still closed at the self-wake, then it is assumed that the door is locked and the system can proceed to the $INPROGRESS$ state. The alternative transition is $abortWash$ which has the negated guard $WM\_doorPosition \neq CLOSED$.

In the first version of our refinement, the proof obligations generated for the safety invariant could not be discharged. 
Model checking does indeed show immediately that the safety invariant is violated and provides a counterexample.
Although the refinement models the latency that  exists between the washing machine sub-system and door sub-system, it allows the user to open and close the door repeatedly in zero-time. Modelling this Zeno behaviour is unrealistic and results in a scenario where the user can close the door and then open it again immediately just before it is locked.

The solution is to model more realistically the latency that must exist in the opening and closing of the door by introducing a delay on the External Event, $closeDoor$. This corresponds to an assumption that the system's time response makes it impossible to open and close the door without it being detected. This is sufficient to ensure that any changes of door state are successfully transmitted to the $WM$ component.
The addition of the latency guard allows the refinement proof obligations to be discharged.

\subsection{The Third Refinement}
Now we refine the notion of the Program. We associate with each $PID$ a $washTime$, $rinseTime$ and $spinTime$ and also introduce a $WashCount$ and $SpinCount$. These properties constrain and make deterministic the operation of the washing machine sub-system for a given $PID$.
The number of washes or rinses associated with a program is modelled using a counter which is decremented and hence completes at $rinseCounter = 0$.

The state machine for this refinement (not shown), invariants concerning the counters have been added to the $INPROGRESS$ and $RINSING$ states. These invariants help ensure that no mistakes have been made in constructing the counters.

\subsection{The Fourth Refinement}

The washing machine sub-system is now further partitioned into two components: the drum sub-system and an abstract component representing the remaining washing machine sub-system, following the pattern of previous refinements.
The component diagram is shown in Figure~\ref{Figure:WM4}.
Three boolean connectors pass messages from the washing machine sub-system to the drum to open and close the hot or cold water valves and to switch the drain pump on or off. Two further natural number connectors pass the water level and the water temperature back from the drum to the washing machine sub-system.
The washing machine state machine is now further refined to manage the filling and emptying of the drum by monitoring the water level as shown.
The value TRUE is sent on the $coldFill$ connector.
The drum sub-system receives the value on the $coldFill$ connector and starts filling the drum.
The drum sub-system sends the value of water level and water temperature repeatedly at unit delay intervals using a self-wake operation $sendLevel$.
The washing machine sub-system switches off the water valve when it detects that the water level associated with the PID has been reached.

\subsection{Summary}
A method for system modelling and refinement has been illustrated using the washing machine case study example. Modelling begins with an abstract state machine model of the system, which is systematically refined into a set of communicating processes: the hardware/software controller under development and the components that represent the controller environment.
At each refinement step, formal proof and model checking are used to validate the model against the requirements and to show absence of deadlock.
The final hardware/software controller component can then be verified within a component-based environment using the CODA Oracle Simulator (Section~\ref{sec:Simulator}).

\section{Refinement of Input and Output}

The previous section demonstrates how to perform abstract modelling and refinement using the CODA components. This section outlines how a refinement can be introduced which models the hardware I/O level behaviour. The model uses a synchronised state-machine, which allows a sequence of clock synchronised I/O events to be performed in order to achieve an abstract data transmission.

\begin{figure}[!htb]
  \centering
  \includegraphics[width=12cm]{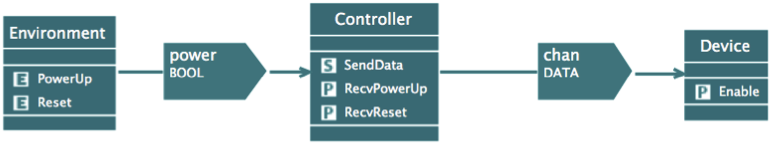}
  \caption{Simple Controller Send Message to Device}
  \label{Figure:io1}
\end{figure}

In the abstract model, shown in Figure~\ref{Figure:io1}, a simple $Controller$ component sends data to enable a $Device$ component. This is modelled as a port-send belonging to a self-wake operation $SendData$, a connector $chan$ and a port-wake operation $Enable$. The self-wake is scheduled by the $RecvPowerUp$ operation, so that its timing represents the completion of an envisaged concrete operation which takes time to complete.

\begin{figure}[!htb]
  \centering
  \includegraphics[width=12cm]{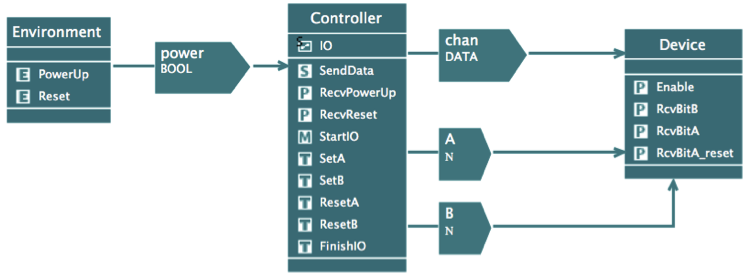}
  \caption{Simple Controller Send Message to Device}
  \label{Figure:io2}
\end{figure}

In the refined model, shown in Figure~\ref{Figure:io2}, the concrete data transmission operations are introduced. In this example, two connectors are used, $A$ for the data bit stream and $B$ for a data ready semaphore. Operations, $SetA$, $SetB$, $ResetA$ and $ResetB$ send 1, 1, 0 and 0 on these connector channels respectively. In order to ensure these operations are invoked in the desired iterative sequence, a synchronised state-machine $IO$ is attached to the $Controller$ component.

\begin{figure}[!htb]
  \centering
  \includegraphics[width=5cm]{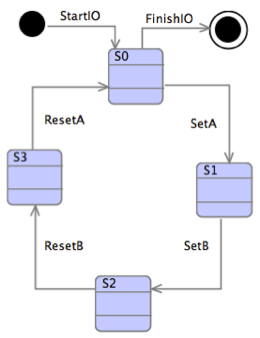}
  \caption{I/O Level State Machine}
  \label{Figure:io3}
\end{figure}

The transitions of this state-machine, Figure~\ref{Figure:io3}, are linked to the same events as the component operations that send the data bits. Hence these operations are constrained to execute in the iterative sequence defined by the state-machine. Furthermore, because this is a synchronised state-machine, it is forced to fire exactly one transition on each clock cycle while it is enabled. The state-machine becomes enabled when the initial transition is taken and this is linked to a method operation of the Controller component that is called by the $RecvPowerUp$ operation. Guards and Actions in the bit sending operations allow the state-machine to complete 16 cyclic sequences before taking the $FinishIO$ transition that disables the state-machine. The data sent by the operation $SendData$ was calculated to be received by operation Enable at the same clock cycle as the last bit is received on connector $A$ representing the end of the bit level transmission.

In this refinement the abstract data connector and associated operation behaviour have been retained although they are now redundant because they are replaced by the I/O level connector behaviour. The abstract data connector can be removed so that we prove that the I/O level is a refinement of the abstract connector.  It may be useful to retain the abstract data connector for later generation of temporal assertions in generated output such as VHDL.

\section{Model checking and simulation with CODA}

\subsection{Model Checking in the CODA Method}

Complex, multi-process embedded systems with inter-process communication are prone to deadlock.  Traditionally in hardware design, state machine abstraction has been shown to help reduce the opportunities for introducing deadlock into a design, but when multiple state machines interact it is possible for one of them to be in the wrong state to receive an incoming message and the system can deadlock.

During the development of the case studies, it was found that ProB animation and deadlock checking provided a necessary adjunct to formal refinement proof to validate the CODA model.  Inadvertent strengthening of event guards can lead to a correct refinement but results in a model which deadlocks.  Since time in CODA is modelled as an incrementing natural number, it is not possible to perform an exhaustive state space search.  However, using the transition coverage metric provided by the ProB model checker, it is possible to show that deadlock is absent when all transitions have been covered.  Although not a proof it gives confidence in the model development and verification process.

\subsection{The CODA Simulator Oracle} \label{sec:Simulator}

\begin{figure}[!htb]
  \centering
  \includegraphics[width=16cm]{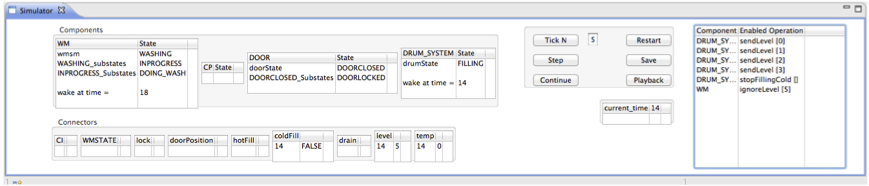}
  \caption{CODA Simulator Display}
  \label{Figure:simulator}
\end{figure}

Our experience with the ProB animator in the CODA case study development highlighted some deficiencies of its standard interface in the CODA flow which have been addressed by developing a simulation  facility within CODA.  
Figure~\ref{Figure:simulator} shows the display for the simulator front-end to ProB that we implemented for CODA.
This has become an important tool within the CODA development and verification method.  First, it operates at the appropriate CODA level of abstraction in terms of Components and Connectors.  Second, it understands the CODA concept of time.  Third it automates the animation process so that simulation tests may be quickly and efficiently generated by the user.  Fourth, it enables animations to be repeated automatically for regression testing purposes and for results to be compared against a golden file.  It is the long term goal of this simulation work to enable tests developed in CODA to drive Orchestra simulations.

During model development, it is useful to ensure that as changes are made at a given level of refinement, the required behaviour is retained.  The simulator oracle ensures that the traces of the variables in the modified model match that of the golden simulation.  The oracle can also be used to verify the variable trace of a refinement against the model it refines to ensure that inadvertent strengthening of the guards has not introduced unintended behaviour.

\section{Related verification work}

A key concept in CODA is the use of state machines augmented with timed operations.  Knapp et al~\cite{DBLP:conf/ftrtft/KnappMR02} describe an approach to reasoning about timed UML state machines.
They augment UML state machines with a timing annotation: a transition from state $A$ to state $B$ annotated with $after(t)$ occurs $t$ time units after state $A$ has become active.
They then define a mapping scheme from these augmented state machines to Uppaal~\cite{DBLP:conf/hybrid/BengtssonLLPY95}.  This allows them to verify temporal properties of timed state machines.  While the $after$ annotation  provides similar modelling capability to CODA, our approach is based on verification of refinement between models rather than verification of temporal properties.

Our Event-B representation of timing behaviour is similar to the approach taken by Cansell et al~\cite{DBLP:conf/b/CansellMR07}.  An activation variable, relating future events to their activation time, is introduced and progress of the clock is constrained so that events occur at exactly their activation time.

Sarshogh \& Butler~\cite{sarshogh11} introduce three discrete timing annotations to Event-B: \textit{delay}, \textit{deadline} and \textit{expiry}.  These allow upper and lower bounds on the duration between trigger-response event pairings.  They identify refinement patterns that allow abstract timing properties to be refined by more complex timing properties involving the introduction of intermediate events between the abstract trigger-response pairs.

\section{Concluding}

We have introduced the CODA modelling and verification framework that builds on Event-B, Rodin, ProB and UML-B to provide a component-based diagrammatic approach to development of embedded systems that matches well the Orchestra approach.  
Our experience to date with several case studies is that the layered approach using refinement fits well with
a development in which requirements are clearly allocated to appropriate refinement levels in a clear and traceable way.  The Event-B basis provides a formal notion of consistency between abstraction levels
of CODA models and Rodin provides the means to verify this consistency. In addition ProB provides a powerful means to validate the accuracy of models through simulation.

\bibliographystyle{eptcs}
\bibliography{coda}
\end{document}